\documentclass[12pt]{article}
\usepackage{amssymb,amsmath,amsthm}
\usepackage{array,longtable}
\usepackage[dvips]{graphicx}
\usepackage[dvips]{color}
\usepackage[mathscr]{eucal}

\newcolumntype{V}{>{$}m{4cm}<{$}}
\newcolumntype{C}{>{$}c<{$}}
\newcolumntype{L}{>{$}l<{$}}
\newcolumntype{R}{>{$}r<{$}}

\newcommand{\Nc}{\mathcal{N}}

\allowdisplaybreaks[3]

\renewcommand{\theequation}{\arabic{section}.\arabic{equation}}

\begin{document}
\title{
$~$\\
\textsc{Open Superstring Star as a Continuous}
\\
\textsc{Moyal Product}
$~$\\
$~$\\}
\author{
\textsf{I.Ya.~Aref'eva and A.A.~Giryavets}
\vspace{5mm}
\\
Steklov Mathematical Institute,
\\
Gubkin st. 8,\\
Moscow, 119991, Russia
\vspace{1mm}
\\
\texttt{arefeva@mi.ras.ru}
\vspace{6mm}
\\
Department of Physics,\\
Moscow State University,\\
Moscow, 119899, Russia
\vspace{1mm}
\\
\texttt{alexgir@mail.ru} }

\date{~}
\maketitle
\thispagestyle{empty}

\begin{abstract}
By  diagonalizing  the three-string vertex
and using  a special coordinate representation the matter part of
the open superstring star is identified with the
continuous Moyal product of functions of anti-commuting variables.
We show that in this representation the identity and sliver
have simple expressions. The relation with the half-string fermionic
variables in continuous basis is given.
\end{abstract}

\newpage
\tableofcontents

\section{Introduction}
\label{sec:intro}
\setcounter{equation}{0}

During the last year it has been realized that
the  vacuum string field theory (VSFT) equations of motion
\cite{0012251}-\cite{0111129} are similar  to the non-commutative soliton
equations  in the large non-commutativity limit \cite{GMS}.
Soliton equations  in this limit are just  projector-like equations.
Solutions of these equations are obtained immediately using an identification
of fields with symbols of operators so that a non-local Moyal product
of two fields  corresponds to an operator product.
To find solutions of VSFT matter equation of motion
\begin{gather*}
|\Phi\rangle*|\Phi\rangle=|\Phi\rangle,
\end{gather*}
where $*$ is the Witten star product \cite{Witten},
it is also useful to identify the string field functionals
with the symbols of operators or with matrices (see \cite{ABGKM} for a review).
Half-string formalism is rather suitable for this purpose
\cite{0105058},\cite{0105059}.
The Witten $*$ and Moyal $*$ were first
identified by Bars using the half-string formalism
\cite{0106157},\cite{0202030}.

Recently, it has been shown by Douglas, Liu, Moore
and Zwiebach \cite{0202087} that the open string star product
in the zero momentum sector
can be described as a continuous tensor product of
mutually commuting two-dimensional Moyal star products.
Non-zero momentum case has been studied in \cite{0204164}.
A continuous parameter $\kappa$
parameterizing components of the continuous tensor product
is nothing but a parameter specifying the eigenvalues of the Neumann
matrices of the three-string vertex operator.
The three-string Neumann matrices
can be diagonalized \cite{0111281}, \cite{0201015}
and there is a basis of oscillators \cite{0202087}
in which the three-string vertex has a simple form
\begin{multline}
|V\rangle_{123}=\exp\Bigl[-\int_{0}^{\infty}\,d\kappa\;
(\tfrac12\mu(\kappa)(o_{\kappa}^{1\,\dagger}o_{\kappa}^{1\,\dagger}
+e_{\kappa}^{1\,\dagger}e_{\kappa}^{1\,\dagger})
+ \mu_s(\kappa)(o_{\kappa}^{1\,\dagger}o_{\kappa}^{2\,\dagger}
+e_{\kappa}^{1\,\dagger} e_{\kappa}^{2\,\dagger})\\
+i\mu_{a}(\kappa)(e_{\kappa}^{1\,\dagger}o_{\kappa}^{2\,\dagger}
-o_{\kappa}^{1\,\dagger} e_{\kappa}^{2\,\dagger})
+ \text{cyc. per.})\Bigr]|0\rangle_{123}.
\label{DLMZ}
\end{multline}
A crucial observation \cite{0202087} is that each factor in \eqref{DLMZ}
is a two-dimensional Moyal product in the oscillator representation
with a parameter of noncommutativity depending on $\kappa$.
Indeed, the Moyal product with a
parameter of noncommutativity $\theta$ of functions
$f$ and $g$ of two-dimensional variables $x$
\begin{equation*}
(f * g)(x^{3}) \equiv \int dx^{1} dx^{2} \;
K(x^{1},x^{2},x^{3})\, f(x^{1}) g(x^{2}),
\end{equation*}
can be rewritten in the oscillator basic as
$|f*g\rangle_3={}_{1}\langle f|{}_{2}\langle g|V^{\text{M}}_{3}\rangle_{123}$,
with $|V^{\text{M}}_3\rangle_{123}$ given by
\begin{multline}
|V^{\text{M}}_3\rangle_{123}=\exp\Bigl[ -\tfrac12 m(\theta)
(o^{1\,\dag}o^{1\,\dag}+e^{1\,\dag}e^{1\,\dag})
-m_s(\theta)(o^{1\,\dag}o^{2\,\dag}+e^{1\,\dag}e^{2\,\dag})\\
-im_a(\theta)(o^{1\,\dag}e^{2\,\dag}-e^{1\,\dag}o^{2\,\dag})
+\hbox{cyc. per.}\Bigr] |0\rangle_{123}.
\label{MV}
\end{multline}
Here $m(\theta)$, $m_s(\theta)$, and $m_a(\theta)$ are given
functions of the parameter of noncommutativity
and $a^i=(e^i,o^i)$ are oscillators corresponding to two-dimensional
Moyal coordinates $x^i=(q^i,p^i)$ ($i=1,2,3$).
One gets this representation using a standard formula for
an eigenvector of the operator
of coordinate in the oscillator representation
\begin{equation}
\label{cor}
\langle x|=\langle 0| \exp [ - \tfrac12 x \cdot
x + i\sqrt{2}\, a \cdot x + \tfrac12 a\cdot a].
\end{equation}
In \cite{0202087} it has been
observed that the components of the three-string vertex
\eqref{DLMZ} have the form
\eqref{MV} for each $\kappa$,
since $\mu(\kappa(\theta))=m(\theta)$,
$\mu_s(\kappa(\theta))=m_s(\theta)$ and $\mu_a(\kappa(\theta))=m_a(\theta)$
for $\theta(\kappa)=2\tanh (\frac{\pi\kappa}{4})$.\\

The purpose of the present paper is a generalization of this result to
the case of the matter sector of the Neveu-Schwarz  superstring.
The Neveu-Schwarz superstring star algebra and its projectors
have been studied in \cite{0112214}-\cite{0204138}.
The spectroscopy problem of Neumann matrices has been solved by Marino and
Schiappa \cite{0112231}, see also appendix of \cite{0203227} for a discussion
of orthonormality and completeness of Neumann matrices eigenvectors.
Note, that it is reasonable
from the very beginning to expect a modification of an analog of
formula \eqref{cor} specifying a suitable coordinate representation
for definition of  the star product as a Moyal product, since supersting
overlaps have a more complicated form in comparison with the
bosonic string overlaps (due to half-integer weights).\\

We begin in section 2 by reminding  the Moyal product
for the functions of anti-commuting variables (see for example
\cite{0002084}, \cite{Berezin}). In section 3 we diagonalize the
three-string matter vertex. In section 4
we identify the matter part of
open superstring star with a continuous Moyal product.
An essential difference from the bosonic case is a
non-trivial dependence of the coordinate representation on
$\kappa$ parameter. Namely, the coordinate representation is
\begin{gather}
\langle \eta|\equiv\langle 0|\exp\Bigl[\int_{0}^{\infty}d\kappa\;
(\frac14 \tilde{f}(\kappa)\eta_{\kappa,\alpha}\,\epsilon_{\alpha\beta}\,
\eta_{\kappa,\beta}
+\eta_{\kappa,\alpha}\,c_{\alpha\beta}\,
\psi_{\kappa,\beta}
+\frac12\,\tau(\kappa)\psi_{\kappa,\alpha}\,\epsilon_{\alpha\beta}\,
\psi_{\kappa,\beta}
)\Bigr],
\end{gather}
where $\eta_{\kappa,\alpha}=(\eta_{\kappa,e},\eta_{\kappa,o})$ and
$\psi_{\kappa,\alpha}=(e_{\kappa},o_{\kappa})$ are two-dimensional
vectors of anti-commuting variables, $\tilde{f}(\kappa)$
and $\tau(\kappa)$ are functions of $\kappa$
and $\epsilon_{\alpha\beta}$ and $c_{\alpha\beta}$
are defined in \eqref{matrices-e} and \eqref{matr-eps-c}. In section 5 we
rewrite the Neveu-Schwarz matter identity and sliver in this
coordinate representation. In section 6 we give a relationship
between the Moyal basis and the half-string fermionic variables.
In section 7 we finish by giving an identification of the Moyal
structures for some special Neveu-Schwarz superghost vertices.

\section{Moyal product of functions of anti-commuting variables}
\label{sec:mpfer}
\setcounter{equation}{0}

Here we describe the Moyal product for functions
of anti-commuting variables $r$ and $s$, $\{r,s\}=0$.
Let us consider a pair of operators $\hat{q}$, $\hat{p}$,
satisfying the anti-commutation relations
\begin{gather}
\{\hat{q},\hat{p}\}=\theta,\quad \{\hat{q},\hat{q}\}=0, \quad
\{\hat{p},\hat{p}\}=0,
\end{gather}
and the Weyl operators depending on the
anti-commuting variables $r$ and $s$
\begin{gather}
U(r,s)=\exp(ir\hat{q}+is\hat{p}).
\end{gather}
These operators satisfy the identity
\begin{equation}
U(r^{1},s^{1})U(r^{2},s^{2}) =U(r^{1}+r^{2},s^{1}+s^{2})
\exp\Bigl[\frac{\theta}{2}(s^{1}r^{2}+r^{1}s^{2})\Bigr].
\end{equation}
If a function $f(q,p)$ of two anti-commuting variables $q$, $p$ is given
in terms of its Fourier transform
\begin{equation}
f(q,p)=\int dsdr\; \exp(-iqr-ips)\tilde{f}(r,s),
\end{equation}
then one can associate it with an operator $\hat{f}$  by the
following formula
\begin{equation}
\hat{f}=\int dsdr\; U(r,s)\tilde{f}(r,s).
\end{equation}

This procedure represents  the Weyl quantization and the function $f(q,p)$
is  the  symbol of the operator $\hat{f}$
(see for example \cite{Berezin},\cite{9907114}).
One has a correspondence
\begin{equation}
\hat{f} \longleftrightarrow f=f(q,p)
\end{equation}
If two operators $\hat{f}_1$ and $\hat{f}_2$ are given by symbols
$f_1(q,p)$ and $f_2(q,p)$
then the symbol of product $\hat{f}_1\hat{f}_2$ is given by the
Moyal product $(f_1* f_2)(q,p)$.
One has
\begin{multline}
\hat{f}_{1}(\hat{p},\hat{q})\hat{f}_{2}(\hat{p},\hat{q})
=\int ds^{1}dr^{1}ds^{2}dr^{2}\;
U(r^{1},s^{1})U(r^{2},s^{2})\tilde{f}_{1}(r^{1},s^{1})
\tilde{f}_{2}(r^{2},s^{2})\\
\equiv\int ds^{3}dr^{3}\; U(r^{3},s^{3})(\widetilde{f_{1}*f_{2}})
(r^{3},s^{3}).
\label{opermultiply}
\end{multline}
Define $k=(r,s)$,  $dk=dsdr$ and
$\Theta_{\alpha\beta}=\begin{pmatrix}
  0 & \theta\\
  \theta & 0
\end{pmatrix}$.
In terms of these notations one finds
\begin{gather}
\label{ff}
(\widetilde{f_{1}*f_{2}})(k^3)
=\int dk^4\;
\exp\Bigl[\frac{1}{2}\Theta_{\alpha\beta}\,k^{4}_{\alpha}k^{3}_{\beta}\Bigr]
\tilde{f}_{1}(k^4+\frac12 k^3)\tilde{f}_{2}(-k^4+\frac12 k^3).
\end{gather}
Introducing notations  $x=(q,p)$ and  $dx=dpdq$ and taking into account that
\begin{gather}
\tilde{f}(k)=\int dx\; \exp(ixk)f(x),\qquad
f(x)=\int dk\; \exp(-ixk)\tilde{f}(k),
\end{gather}
one can rewrite  (\ref{ff}) in terms of Fourier components
\begin{gather}
(f_{1}*f_{2})(x^{3})=\int dk^3\; \exp(-ix^{3}k^3)(\widetilde{f_{1}*f_{2}})(k^3)
\equiv\int dx^{1}dx^{2}\;K(x^{1},x^{2},x^{3})f_{1}(x^{1})f_{2}(x^{2}).
\label{moyal-product}
\end{gather}
Integrating one gets the following expression for
the kernel K:
\begin{gather}
K(x^{1},x^{2},x^{3})
=\frac{\theta^{2}}{4}
\exp[2(x^{2}-x^{3})_{\alpha}\,\Theta^{-1}_{\alpha\beta}\,(x^{3}-x^{1})_{\beta}].
\label{kernel}
\end{gather}
Notice a similarity of this formula with a corresponding
bosonic formula \cite{0202087}.
One finds
\begin{gather}
\{q,p\}_{*}=\theta.
\end{gather}
Also one has $\lim _{\theta \to 0} f*g=fg$.

\section{Open superstring star in the continuous oscillator basis}
\label{sec:mpver}
\setcounter{equation}{0}
\subsection{Review of spectroscopy of the vertices}

The commuting real symmetric matrices $F_{rs}$ and $(C\tilde{F})_{rs}$
(twist matrix $C_{rs}=(-1)^{r+\tfrac12}\delta_{rs}$, $r,s\geq \frac12$)
that specify
the Neveu-Schwarz vertices have the common set of eigenvectors
$v_{s}(\kappa)$ which
are labelled by a continuous parameter $\kappa\in(-\infty,\infty)$,
\begin{gather}
\sum_{s\geq\tfrac12}F_{rs}v_{s}(\kappa)
=f(\kappa)v_{r}(\kappa),\qquad
\sum_{s\geq\tfrac12}(C\tilde{F})_{rs}v_{s}(\kappa)
=\tilde{f}(\kappa)v_{r}(\kappa).
\end{gather}
and are given by the generating
function \cite{0112231}
\begin{gather}
f_{\kappa}(z)
=\sum_{r\geq\tfrac12}v_{r}(\kappa)z^{r+\tfrac12}
=\mathcal{N}(\kappa)^{-\tfrac12}
\frac{z}{\sqrt{1+z^{2}}}\exp(-\kappa\arctan(z)),
\end{gather}
with  normalization factor
$\mathcal{N}(\kappa)=2\cosh\left(\frac{\pi\kappa}{2}\right)$ \cite{0203227}.
The eigenvalues are given by \cite{0112231}
\begin{gather}
f(\kappa)=-\frac{1}{\cosh(\frac{\pi\kappa}{2})}
=-\frac{1-\tau^2(\kappa)}{1+\tau^2(\kappa)},\qquad
\tilde{f}(\kappa)=-\tanh\left(\frac{\pi\kappa}{2}\right)
=-\frac{2\tau(\kappa)}{1+\tau^2(\kappa)},
\end{gather}
where $\tau(\kappa)=\tanh(\frac{\pi\kappa}{4})$.
The twist matrix $C_{rs}$ acts on eigenvectors as follows
\begin{gather}
\sum_{s\geq\tfrac12}C_{rs}v_{s}(\kappa)=-v_{r}(-\kappa).
\end{gather}
and therefore odd and even components of
the eigenvectors satisfy the relations
\begin{gather}
v_{r_o}(-\kappa)=v_{r_o}(\kappa),\qquad
v_{r_e}(-\kappa)=-v_{r_e}(\kappa).\label{even-odd}
\end{gather}
Here we defined even $r_{e}=2n-\tfrac12$ and odd $r_{o}=2n-\tfrac32$
indices ($n\geq1$).
The eigenvectors $v_{r}(\kappa)$ are orthogonal and complete \cite{0203227}
\begin{gather}
\sum_{r\geq\tfrac12}v_{r}(\kappa_{1})v_{r}(\kappa_{2})
=\delta(\kappa_{1}-\kappa_{2}),\qquad
\int_{-\infty}^{\infty}d\kappa\;
v_{r}(\kappa)v_{s}(\kappa)=\delta_{rs}.
\label{comp-ortog}
\end{gather}

\subsection{Diagonalization of the vertex}

As in the bosonic case \cite{0202087},\cite{0204164},
instead of usual Neveu-Schwarz matter oscillators
($\psi_{r},\psi^{\dag}_{r}=\psi_{-r}$, $r\geq\tfrac12$)
it is convenient to introduce the
continuous oscillators (Lorentz index is omitted)
\begin{gather}
\tilde{\psi}_\kappa
=\sum_{r\geq\tfrac12}v_{r}(\kappa)\psi_{r},\qquad
\tilde{\psi}_\kappa^{\dag}=\sum_{r\geq\tfrac12}v_{r}(\kappa)\psi_{r}^{\dag}.
\label{cont-psi-osc}
\end{gather}
Due to anti-commutation relations $\{\psi_{r},\psi^{\dag}_{s}\}=\delta_{rs}$
the continuous oscillators satisfy the following anti-commutation relation
\begin{gather}
\{\,\tilde{\psi}_\kappa,\tilde{\psi}_{\kappa'}^\dagger \}=\delta(\kappa - \kappa').
\label{anticommut-rel}
\end{gather}
The inverse of relations \eqref{cont-psi-osc} are
\begin{gather}
\psi_{r}
=\int_{-\infty}^{\infty}d\kappa\,v_{r}(\kappa)\tilde{\psi}_{\kappa},\qquad
\psi^{\dag}_{r}
=\int_{-\infty}^{\infty}d\kappa\,v_{r}(\kappa)\tilde{\psi}^{\dag}_{\kappa}.
\label{osc-ful-diag}
\end{gather}
Using the equations \eqref{eigen-eq}, \eqref{osc-ful-diag} and the completeness
relations \eqref{comp-ortog} the three-string vertex \cite{GJ3}
\begin{gather}
|V_3\rangle_{123}
=\exp\Bigl[\frac12\sum_{a,b=1}^{3}\sum_{r,s\geq\tfrac12}
\psi_{r}^{a\,\dagger}(CM^{ab})_{rs}
\psi_{s}^{b\,\dagger}\Bigr]|0\rangle_{123}
\label{vert-psi}
\end{gather}
can be rewritten in the continuous basis
\begin{gather}
|V_3\rangle_{123}=\exp\Bigl[\frac12\sum_{a,b=1}^{3}
\int_{-\infty}^{\infty}d\kappa\,
\mu^{ab}(\kappa)C\tilde{\psi}^{\dag\,a}_{\kappa}\,
\tilde{\psi}^{\dag\,b}_{\kappa}
\Bigr]|0\rangle_{123}.
\label{diag-psi}
\end{gather}
Here the twist operator $C$ acts in the continuous basis as follows:
$C\tilde{\psi}^{\dag}_{\kappa}=-\tilde{\psi}^{\dag}_{-\kappa}$.
Let us introduce even and odd  continuous
oscillators $e^{\dag}_\kappa$ and $o^{\dag}_\kappa$
with respect to $C$ conjugation
\begin{subequations}
\begin{align}
&e_\kappa^\dagger
=\frac{1}{\sqrt{2}}(\tilde{\psi}^{\dag}_{\kappa}+C\tilde{\psi}^{\dag}_{\kappa})
=\sqrt{2}\sum_{r_e}v_{r_e}(\kappa)\,\psi_{r_e}^\dagger,\qquad
&\psi_{r_e}^{\dagger}&
=\sqrt{2}\int_{0}^{\infty}d\kappa\,v_{r_e}(\kappa)\,e_{\kappa}^{\dagger},\\
&o_\kappa^\dagger
=\frac{1}{\sqrt{2}}(\tilde{\psi}^{\dag}_{\kappa}-C\tilde{\psi}^{\dag}_{\kappa})
=\sqrt{2}\sum_{r_o}v_{r_o}(\kappa)\,\psi_{r_o}^\dagger,\qquad
&\psi_{r_o}^{\dagger}&
=\sqrt{2}\int_{0}^{\infty}d\kappa\,v_{r_o}(\kappa)\,o_{\kappa}^{\dagger}.
\end{align}
\label{osc-def}
\end{subequations}
In this basis the twist $C$ matrix has a diagonal form
\begin{gather}
c_{\alpha\beta}=\begin{pmatrix}
  1 & 0 \\
  0 & -1
\end{pmatrix}.
\label{matr-eps-c}
\end{gather}
Define the two dimensional variable
$\psi^{\dag}_{\kappa,\alpha}=(e^{\dag}_{\kappa},o^{\dag}_{\kappa})$.
In these notations
the anti-commutation relation \eqref{anticommut-rel}
takes the form
\begin{gather}
\{\,\psi_{\kappa,\alpha},\psi_{\kappa',\beta}^\dagger \}
=\delta_{\alpha\beta}\delta(\kappa - \kappa'),\qquad
\kappa, \kappa'>0.
\end{gather}
Note that $C\psi_{\kappa,\alpha}=c_{\alpha\beta}\psi_{\kappa,\beta}$.
The BPZ conjugation of $\psi_{-r}$ and $\psi_{r}$ are given by (bpz$^2=1$)
\begin{gather}
\text{bpz}(\psi_{-r})=(-1)^{-r+\tfrac12}\psi_{r},\qquad
\text{bpz}(\psi_{r})=(-1)^{r-\tfrac12}\psi_{-r},\qquad r\geq\tfrac12.
\label{bpz-conj-psi}
\end{gather}
The BPZ conjugation does not change ordering
$\text{bpz}(\psi_{-r}\psi_{-s})=\text{bpz}(\psi_{-r})\text{bpz}(\psi_{-s})$.
Using \eqref{bpz-conj-psi}
one finds the following BPZ conjugation of the continuous oscillators
\begin{gather}
\text{bpz}(\psi^{\dag}_{\kappa,\alpha})=-c_{\alpha\beta}\psi_{\kappa,\beta},\qquad
\text{bpz}(\psi_{\kappa,\alpha})=-c_{\alpha\beta}\psi^{\dag}_{\kappa,\beta}.
\label{bpz-psi-eo}
\end{gather}
Using the relations
$e_{-\kappa}^\dagger=-e_{\kappa}^\dagger$ and
$o_{-\kappa}^\dagger=o_{\kappa}^\dagger$
one can rewrite
the three-string vertex \eqref{diag-psi} in
terms of oscillators
$\psi^{\dag}_{\kappa,\alpha}=(e_{\kappa}^{\dag}, o_{\kappa}^{\dag})$ as
\begin{gather}
|V_3\rangle_{123}
=\exp\Bigl[\frac12 \int_{0}^{\infty}d\kappa\;
\psi^{a\,\dag}_{\kappa,\alpha}V^{ab}_{\kappa,\alpha\beta}
\psi^{b\,\dag}_{\kappa,\beta}\Bigr]
|0\rangle_{123},
\label{vertex-diag}
\end{gather}
where $V^{ab}_{\kappa,\alpha\beta}$ is $6\times 6$
matrix defined by
\begin{gather}
V^{ab}_{\kappa,\alpha\beta}=
\mu(\kappa)\,\epsilon_{\alpha\beta}\otimes\delta^{ab}
+\mu_a(\kappa)\,c_{\alpha\beta}\otimes\chi^{ab}
+\mu_s(\kappa)\,\epsilon_{\alpha\beta}\otimes\varepsilon^{ab};
\end{gather}
\begin{gather}
\mu\equiv\mu^{11}=\tau\frac{1-\tau^2}{1+3\tau^2},\quad
\mu_{s}\equiv\frac12(\mu^{12}+\mu^{21})=
-\tau\frac{1+\tau^2}{1+3\tau^2},\quad
\mu_a\equiv\frac12(\mu^{12}-\mu^{21})
=\frac{1+\tau^2}{1+3\tau^2}.
\label{mu-tau}
\end{gather}
Also we use the notations of \cite{0207174}
\begin{gather}
\chi^{ab}=\begin{pmatrix}
  0 & 1 & -1 \\
  -1 & 0 & 1 \\
  1 & -1 & 0
\end{pmatrix},\qquad
\varepsilon^{ab}=\begin{pmatrix}
  0 & 1 & 1 \\
  1 & 0 & 1 \\
  1 & 1 & 0
\end{pmatrix},\qquad
\epsilon_{\alpha\beta}=\begin{pmatrix}
  0 & 1 \\
  -1 & 0
\end{pmatrix}.
\label{matrices-e}
\end{gather}
Here and further the summation over $a,b=1,2,3$ is assumed.

\subsection{Diagonalization of the identity and sliver}

For the identity and sliver one finds the following expressions
in the continuous basis
\begin{align}
&|I\rangle
=\exp\Bigl[\frac12\sum_{r,s\geq\tfrac12}\psi_{r}^{\dagger}
I_{rs}\psi_{s}^{\dagger}\Bigr]|0\rangle
=\exp\Bigl[-\int_{0}^{\infty}d\kappa\;
\tau(\kappa) e_{\kappa}^{\dag}o_{\kappa}^{\dag}\Bigr]|0\rangle,
\label{diag-iden}
\\
&|\Xi\rangle
=\exp\Bigl[\frac12\sum_{r,s\geq\tfrac12}
\psi_{r}^{\dagger}
(CT)_{rs}
\psi_{s}^{\dagger}\Bigr]|0\rangle
=\exp\Bigl[\int_{0}^{\infty}d\kappa\;
T(\kappa) e_{\kappa}^{\dag}o_{\kappa}^{\dag}\Bigr]
|0\rangle,
\label{diag-sliv}
\end{align}
where $T(\kappa)=\exp(-\tfrac12 \pi\kappa)
=\dfrac{1-\tau(\kappa)}{1+\tau(\kappa)}$.
$\Nc_{I}$ and $\Nc_{\Xi}$ are defined to be the norms
of the surface states
\begin{gather}
\Nc_{I}\equiv\langle I|I\rangle=(\det(1+(CI)^2))^{\tfrac12}\quad
\text{and}\quad
\Nc_{\Xi}\equiv\langle \Xi|\Xi\rangle=(\det(1+T^2))^{\tfrac12}.
\end{gather}
Actually these norms are infinite
and the inclusion of ghost sector
will apparently cancel the singularities.
A similar problem in the bosonic case is discussed
for example in \cite{0207174}.

\section{Identification of Moyal structures}
\label{sec:mpiden}
\setcounter{equation}{0}

In this section we identify the Moyal structures in the
three-string vertex by finding an appropriate coordinate representation.
In the first subsection we introduce the coordinate representation
and find the partition of unity.
In the second subsection we use this coordinate representation
for identification of the three-string star with the Moyal product
of functions of anti-commuting variables introduced in section 2.

\subsection{Coordinate representation}

Introduce the following continuous coordinate representation
\begin{gather}
\langle \eta|\equiv\langle 0|\exp\Bigl[\int_{0}^{\infty}d\kappa\;
(\frac14 \tilde{f}(\kappa)\eta_{\kappa,\alpha}\,\epsilon_{\alpha\beta}\,
\eta_{\kappa,\beta}
+\eta_{\kappa,\alpha}\,c_{\alpha\beta}\,
\psi_{\kappa,\beta}
+\frac12\,\tau(\kappa)\psi_{\kappa,\alpha}\,\epsilon_{\alpha\beta}\,
\psi_{\kappa,\beta}
)\Bigr],\label{coor-eta}
\end{gather}
where $\eta_{\kappa,\alpha}=(\eta_{\kappa,e},\eta_{\kappa,o})$
are anti-commuting variables.
Actually these variables can be rescaled.
We choose a fixed scale. The BPZ
conjugated state $\text{bpz}(\langle\eta|)\equiv|\eta\rangle$ is given
by (use \eqref{bpz-psi-eo})
\begin{gather}
|\eta\rangle=\exp\Bigl[ \int_{0}^{\infty}d\kappa\;
(\frac14 \tilde{f}(\kappa)\eta_{\kappa,\alpha}\,\epsilon_{\alpha\beta}\,
\eta_{\kappa,\beta}
-\eta_{\kappa,\alpha}\,\delta_{\alpha\beta}\,
\psi^{\dag}_{\kappa,\beta}
-\frac12\,\tau(\kappa)\psi_{\kappa,\alpha}^{\dag}\,\epsilon_{\alpha\beta}\,
\psi_{\kappa,\beta}^{\dag})
\Bigr]|0\rangle.
\label{coor-repr-right}
\end{gather}
The states $\langle\eta|$ are eigenstates of the operators
\begin{subequations}
\begin{gather}
\hat{\eta}_{\kappa,e}
=e^{\dag}_{\kappa}+\tau(\kappa)o_{\kappa},\qquad
\hat{\eta}_{\kappa,o}
=\tau(\kappa) e_{\kappa}-o^{\dag}_{\kappa},
\end{gather}
\end{subequations}
with eigenvalues $\eta_{\kappa,e}$ and $\eta_{\kappa,o}$. Also
\begin{gather}
\{\hat{\eta}_{\kappa,\alpha},\hat{\eta}^{\dag}_{\kappa',\beta}\}
=\theta(\kappa)\delta_{\alpha\beta}\,\delta(\kappa-\kappa'),
\end{gather}
where $\theta(\kappa)=1+\tau^{2}(\kappa)$.
To understand a meaning of the coordinates $\eta_{\kappa,\alpha}$
it is useful to rewrite them in the discrete basis:
\begin{subequations}
\begin{align}
\tilde{\eta}_{\kappa}&=\sum_{r\geq\tfrac12}v_{r}(\kappa)\eta_{r},\\
\eta_{\kappa,e}&=\frac{1}{\sqrt{2}}(\tilde{\eta}_{\kappa}+C\tilde{\eta}_{\kappa})
=\sqrt{2}\sum_{r_e}v_{r_e}(\kappa)\eta_{r_e},\\
\eta_{\kappa,o}&=\frac{1}{\sqrt{2}}(\tilde{\eta}_{\kappa}-C\tilde{\eta}_{\kappa})
=\sqrt{2}\sum_{r_o}v_{r_o}(\kappa)\eta_{r_o}.
\end{align}
\end{subequations}
In terms of these discrete variables the coordinate representation
\eqref{coor-repr-right} takes the form
\begin{gather}
|\eta\rangle=\exp\Bigl[
\,\frac14\,\sum_{r,s\geq\tfrac12}\eta_r\,\tilde{F}_{rs}\,\eta_s
-\sum_{r\geq\tfrac12}\eta_r\,\psi^{\dag}_r
+\frac12\,\sum_{r,s\geq\tfrac12}\psi^{\dag}_r\,I_{rs}\,\psi^{\dag}_s
\Bigr]|0\rangle.
\label{coor-repr-right-discr}
\end{gather}
One sees that $|\eta\rangle$ has an interpretation
of a coherent state over the identity surface state.
One finds
\begin{gather}
\langle \eta|\lambda\rangle
=\Nc_{I}\exp\Bigl[
\int_{0}^{\infty}d\kappa\;
\frac{1}{\theta(\kappa)}\,\eta_{\kappa,\alpha}\,
c_{\alpha\beta}\,\lambda_{\kappa,\beta}\Bigr]
=\Nc_{I}\exp\Bigl[
\,\frac12\,\sum_{r,s\geq\tfrac12}\eta_{r}\,((1-F)C)_{rs}\,\lambda_s\Bigr].
\label{norm-equation}
\end{gather}
This provides the following
partition of unit operator in continuous and discrete basis:
\begin{subequations}
\begin{gather}
1=\Nc_{I}\int
\mathbf{d}\eta\,\mathbf{d}\lambda\,
\exp\Bigl[
\int_{0}^{\infty}d\kappa\;
\frac{1}{\theta(\kappa)}\, \eta_{\kappa,\alpha}\,c_{\alpha\beta}\,
\lambda_{\kappa,\beta}\Bigr]
|\eta\rangle\langle \lambda|;\\
1=\Nc_{I}\int
\mathbf{d}\eta\,\mathbf{d}\lambda\,
\exp\Bigr[
\,\frac12\sum_{r,s\geq\tfrac12}\eta_{r}\,((1-F)C)_{rs}\,\lambda_s\Bigl]
|\eta\rangle\langle \lambda|,
\end{gather}
\label{edinitsa}
\end{subequations}
where
$\mathbf{d}\eta\equiv\prod_{r\geq\tfrac12}
d\eta_{r}$.

\subsection{Open superstring star as a continuous Moyal product}

In this subsection we insert two partitions of unity
\eqref{edinitsa} in the three-string
star product and identify it with the Moyal product
of functions of anti-commuting variables introduced in section 2.
The string field
functional corresponding to string
field $|\Psi\rangle$
is given by
$\Psi(\eta)\equiv\langle\eta|\Psi\rangle=\langle\Psi|\eta\rangle$.
Inserting two partitions of unity
\eqref{edinitsa} in the three-string star product
\begin{gather}
|\Psi^{1}*\Psi^{2}\rangle_{3}
={}_{1}\langle\Psi^1|{}_{2}\langle\Psi^{2}|V_{3}\rangle_{123}
\end{gather}
and multiplying it with $\langle\eta^3|$ one obtains
\begin{multline}
\langle \eta^{3}|\Psi^{1}*\Psi^{2}\rangle=
\Nc_{I}^2
\int
\prod_{a=1,2}\,
\mathbf{d}\eta^{a}\,\mathbf{d}\lambda^{a}\,
\langle\Psi^{a}|\eta^{a}\rangle
\exp\Bigl[\int_{0}^{\infty}d\kappa\;
\frac{1}{\theta(\kappa)}\,\eta^{a}_{\kappa,\alpha}\,
c_{\alpha\beta}\,\lambda^{a}_{\kappa,\beta}\Bigr]\;
K(\lambda^1,\lambda^2,\eta^3)\\
=\int\mathbf{d}\eta^{1}\,\mathbf{d}\eta^{2}\,
\Psi^{1}(\eta^{1})\Psi^{2}(\eta^{2})
\tilde{K}(\eta^1,\eta^2,\eta^3).
\label{ins-two-id}
\end{multline}
Here the kernel $K(\eta^1,\eta^2,\eta^3)$ is given by
\begin{gather}
K(\eta^1,\eta^2,\eta^3)
\equiv
{}_{1}\langle \eta^1|{}_{2}\langle \eta^2|{}_{3}\langle \eta^3|
V_{3}\rangle_{123}
=\Nc_{K}\exp\Bigl[\frac12\,\int_{0}^{\infty}d\kappa\,
\frac{1}{\theta(\kappa)}\,\eta^a_{\kappa,\alpha}\,
c_{\alpha\beta}\otimes\chi^{ab}\,\eta^b_{\kappa,\beta}\Bigr],
\label{kernel-eta}
\end{gather}
where normalization $\Nc_{K}$ is of the form
\begin{gather}
\Nc_{K}=\left[\det\left(
\tfrac14(1-F)^{2}(2+F)\right)\right]^{-\tfrac12}.
\end{gather}
Integration over $\lambda^1$ and $\lambda^2$ in \eqref{ins-two-id} yields
\begin{multline}
\tilde{K}(\eta^1,\eta^2,\eta^3)=
\Nc_{K}\Nc_{I}^2\int
\mathbf{d}\lambda^{1}\,
\mathbf{d}\lambda^{2}\,
\exp\Bigl[\int_{0}^{\infty}d\kappa\;
\frac{1}{\theta(\kappa)}\,
(\sum_{a=1,2}\eta^{a}_{\kappa,\alpha}\,c_{\alpha\beta}\,
\lambda^{a}_{\kappa,\beta})\Bigr]\\
\times
\exp\Bigl[\int_{0}^{\infty}d\kappa\;
\frac{1}{\theta(\kappa)}\,
(\lambda^1_{\kappa,\alpha}\,c_{\alpha\beta}\,\lambda^2_{\kappa,\beta}
+\lambda^2_{\kappa,\alpha}\,c_{\alpha\beta}\,\eta^3_{\kappa,\beta}
+\eta^3_{\kappa,\alpha}\,c_{\alpha\beta}\,\lambda^1_{\kappa,\beta})\Bigr]
=K(-\eta^1,\eta^2,\eta^3).
\label{kernel-tilde}
\end{multline}
Finally one finds
\begin{multline}
(\Psi^{1}*\Psi^{2})(\eta^3)
=
\int \mathbf{d}\eta^{1}\,\mathbf{d}\eta^{2}\,
\Psi^{1}(\eta^{1})
\Psi^{2}(\eta^{2})
K(-\eta^1,\eta^2,\eta^3)\\
=
\int \mathbf{d}\eta^{1}\,\mathbf{d}\eta^{2}\,
\Psi^{1}(-\eta^{1})
\Psi^{2}(\eta^{2})
K(\eta^1,\eta^2,\eta^3).
\label{star-prod-ker-minus}
\end{multline}
Note that there is
a minus sign in the argument of
functional $\Psi^1$ in the last line of \eqref{star-prod-ker-minus}.
This is a mild (up to a sign) associativity anomaly which was found
in \cite{BJM} and has been recently discussed in \cite{0112231}.
Note that this anomaly
exist only in the GSO$-$ sector. The star product is
associative in the GSO$+$ sector.

For the identification
with the Moyal product obtained in section 2
instead of $\eta_{\kappa,\alpha}=(\eta_{\kappa,e},\eta_{\kappa,o})$
it is relevant to use the
anti-commuting variables $x_{\kappa}=(q_{\kappa},p_{\kappa})$
\begin{gather}
q_{\kappa}=\frac{1}{2}(\eta_{\kappa,e}+\eta_{\kappa,o}),\qquad
p_{\kappa}=\frac{1}{2}(\eta_{\kappa,e}-\eta_{\kappa,o}).
\end{gather}
In terms of these variables the kernel \eqref{kernel-eta} takes the form
\begin{gather}
K(x^{1},x^{2},x^{3})\equiv K(\eta^1,\eta^2,\eta^3)
=\exp\Bigl[\int_{0}^{\infty}d\kappa\;\frac{2}{\theta(\kappa)}\,
q^{a}_{\kappa}\,\chi^{ab}\,p^{b}_{\kappa}\Bigr].
\end{gather}
Up to normalization this coincides for every $\kappa$
with the Moyal kernel \eqref{kernel}.
Further we will also use the notation $\langle x|$
instead of
$\langle\eta|$
($\langle x|\equiv\langle\eta|$).
Note also that
\begin{gather}
\{q_{\kappa},p_{\kappa'}\}_{*}=\theta(\kappa)\delta(\kappa-\kappa').
\end{gather}

\section{Sliver, identity and wedge states}
\label{sec:mpstarpr}
\setcounter{equation}{0}

Now we
are going to rewrite the well known star algebra projectors,
the identity \eqref{diag-iden}
and sliver \eqref{diag-sliv},
in the $x_{\kappa}=(p_{\kappa},q_{\kappa})$ basis.
In the coordinate representation the sliver is given by
\begin{equation}
\langle
x|\Xi\rangle=(\det(1+CIT))^{\tfrac12}
\exp\Bigl[\int_{0}^{\infty}d\kappa\;
\frac{2}{\theta(\kappa)}\,q_{\kappa}p_{\kappa}\Bigr].
\end{equation}
One expects this answer since $f(x)=\tfrac{1}{2}\exp(\,\tfrac{2}{\theta}\,qp\,)$
is a projector with respect to the Moyal product of functions of anti-commuting
variables \eqref{moyal-product}.
In the coordinate representation the identity is given by
\begin{gather}
\langle x|I\rangle=\Nc_{I}.
\end{gather}
This is also an expected answer since $f(x)=1$ is a projector of the
Moyal product \eqref{moyal-product}.
So we found that up to normalization
the sliver and identity in the coordinate representation
are the projectors with respect to the Moyal product.

A few comment on wedge states.
It is known that the star algebra has a subalgebra of
the wedge states $|n\rangle$, $n\geq 1$
with the multiplication rule \cite{RZ}
\footnote{For algebraic construction of wedge states
for Neveu-Schwarz string see \cite{0203227}.}
\begin{gather}
|n\rangle*|m\rangle=|n+m-1\rangle,
\end{gather}
where $|1\rangle$ corresponds to the identity $|I\rangle$
and $|2\rangle$ corresponds
to the Neveu-Schwarz superstring vacuum $|0\rangle$.
The wedge states subalgebra
has a natural interpretation in terms of the Moyal
product of wedge states in coordinate representation.
The ground state and the identity
functions are  $f^{1}(x)\equiv\langle x|0\rangle$ and
$f^{0}(x)\equiv\langle x|I\rangle$, respectively.
It is known that the wedge states $|n\rangle$ can be obtained
by multiplying the vacuum $n-1$ times,
i.e. $|n\rangle=(|0\rangle)_{*}^{n-1}$.
In the functional language $f^{n-1}(x)\equiv\langle x|n\rangle$
and one gets a correspondence
\begin{gather}
\underbrace{|0\rangle *|0\rangle* ...*|0\rangle}_{n-1}=|n\rangle \longleftrightarrow
f^{n-1}(x)=\underbrace{f^{1}*f^{1}*...*f^{1}}_{n-1}(x)
\end{gather}
for $n\geq 2$. This means
that there is a subalgebra of algebra with a Moyal product.
Basis elements of this subalgebra are $f^{n}(x)$; $f^{0}(x)$
is the identity and the multiplication rule is given by
\begin{gather}
(f^{n}*f^{m})(x)=f^{n+m}(x).
\end{gather}

\section{Relation with half-string fermionic variables}
\label{sec:half-fermi}
\setcounter{equation}{0}

In this section we give a relationship between the continuous Moyal basis
and the half-string fermionic variables.
To work out half-sting fermionic variables it is more convenient
to use $x_{\kappa}=(q_{\kappa},p_{\kappa})$ basis.
In this basis \eqref{star-prod-ker-minus} takes the form
\begin{multline}
(\Psi^{1}*\Psi^{2})(q^3,p^3)
=\Nc_{K}
\int\mathbf{d}q^{1}\;\mathbf{d}p^{1}\;\mathbf{d}q^{2}\;\mathbf{d}p^{2}\;
\Psi^{1}(-q^{1},-p^{1})
\Psi^{2}(q^{2},p^{2})\exp\Bigl[\int_{0}^{\infty}d\kappa\;
\frac{2}{\theta(\kappa)}\,
q^{a}_{\kappa}\,\chi^{ab}\,p^{b}_{\kappa}\Bigr].
\label{star-prod-x}
\end{multline}
Often for simplicity we will omit index $\kappa$ in the coordinates.
Performing Fourier transformation in $p_{\kappa}$ variable
\begin{gather}
\Psi(q,u)\equiv\int \mathbf{d}p\; \exp\Bigl[\int_{0}^{\infty}d\kappa\;
 p_{\kappa}u_{\kappa}\Bigr]\Psi(q,p),
\end{gather}
where $\mathbf{d}p\equiv\prod_{r\geq\tfrac12}dp_{r}$ for discrete basis and
introducing the notation
\begin{gather}
\Psi(q,u)\equiv\tilde{\Psi}\Bigl(q_{\kappa}-\frac{\theta(\kappa)}{2}u_{\kappa},
q_{\kappa}+\frac{\theta(\kappa)}{2}u_{\kappa}\Bigr)
\equiv\tilde{\Psi}(\psi^{l},\psi^{r})
\end{gather}
one rewrites \eqref{star-prod-x} as
\begin{gather}
(\widetilde{\Psi^{1}*\Psi^{2}})(\psi^{l},\psi^{r})
=\int \mathbf{d}p\; \exp\Bigl[\int_{0}^{\infty}d\kappa\;
 p_{\kappa}u_{\kappa}\Bigr](\Psi^{1}*\Psi^{2})(q,p)
=\Nc_{H}\int \mathbf{d}\xi\;
\tilde{\Psi}^{1}(-\psi^{l},-\xi)
\tilde{\Psi}^{2}(\xi,\psi^{r}),
\label{star-half-ful}
\end{gather}
where $\Nc_{H}$ is normalization constant.
The operators corresponding to the left and right
fermionic variables $\psi^{l}$ and $\psi^{r}$
in the oscillator representation are given by
\begin{subequations}
\begin{gather}
\hat{\psi}^{l}_{\kappa}
=\hat{q}_{\kappa}-\frac{\theta(\kappa)}{2}\hat{u}_{\kappa}=
\frac{1}{\sqrt{2}}((\tau(\kappa)-1)\tilde{\psi}_{\kappa}
+(\tau(\kappa)+1)C\tilde{\psi}_{\kappa}^{\dag}),\\
\hat{\psi}^{r}_{\kappa}
=\hat{q}_{\kappa}+\frac{\theta(\kappa)}{2}\hat{u}_{\kappa}=
\frac{1}{\sqrt{2}}((\tau(\kappa)+1)\tilde{\psi}_{\kappa}
-(\tau(\kappa)-1)C\tilde{\psi}_{\kappa}^{\dag}).
\end{gather}
\label{half-str}
\end{subequations}
Here $\hat{u}_{\kappa}
=\frac{1}{\theta(\kappa)}
(\hat{\eta}^{\dag}_{\kappa,e}-\hat{\eta}^{\dag}_{\kappa,o})$.
In discrere basis \eqref{half-str} takes the form
\begin{subequations}
\begin{gather}
\hat{\psi}^{l}_{\kappa}=-\frac{1}{\sqrt{2}\tilde{f}(\kappa)}
((C\psi^{\dag}+F\psi+\tilde{F}\psi^{\dag})_{s}
+C_{sr}(C\psi+F\psi^{\dag}+\tilde{F}\psi)_{r})v_{s}(\kappa),\\
\hat{\psi}^{r}_{\kappa}=-\frac{1}{\sqrt{2}\tilde{f}(\kappa)}
((-C\psi^{\dag}+F\psi+\tilde{F}\psi^{\dag})_{s}
-C_{sr}(-C\psi+F\psi^{\dag}+\tilde{F}\psi)_{r})v_{s}(\kappa).
\end{gather}
\label{half-str-discr}
\end{subequations}
Formulae \eqref{half-str-discr}
suggest the following half-string fermionic variables
\begin{gather}
l_{r}=\frac{1}{\pi}\int_{-\frac{\pi}{2}}^{\frac{\pi}{2}}d\sigma\;
e^{-ir\sigma}\psi(\sigma),\qquad
r_{s}=\frac{1}{\pi}\int_{-\frac{\pi}{2}}^{\frac{\pi}{2}}d\sigma\;
e^{-is\sigma}i\psi(\pi-\sigma),
\end{gather}
such that
\begin{subequations}
\begin{align}
l_{r_e}&=\psi_{r_e}+F_{r_e\,s_e}\psi^{\dag}_{s_e}+\tilde{F}_{r_e\,s_o}\psi_{s_o},
&l_{r_o}=\psi_{r_o}-F_{r_o\,s_o}\psi^{\dag}_{s_o}-\tilde{F}_{r_o\,s_e}\psi_{s_e};\\
r_{r_e}&=-\psi^{\dag}_{r_e}+F_{r_e\,s_e}\psi_{s_e}+\tilde{F}_{r_e\,s_o}\psi^{\dag}_{s_o},
&r_{r_o}=\psi^{\dag}_{r_o}+F_{r_o\,s_o}\psi_{s_o}+\tilde{F}_{r_o\,s_e}\psi^{\dag}_{s_e}.
\end{align}
\label{left-right-osc}
\end{subequations}
In this case the left operator $\hat{\psi}^{l}_{\kappa}$
and the right operator $\hat{\psi}^{r}_{\kappa}$ are expressed in terms
of the left and right fermionic oscillators \eqref{left-right-osc}
\begin{gather}
\psi^{l}_{\kappa}
=-\sum_{r,s\geq \tfrac12}\frac{1}{\sqrt{2}\tilde{f}(\kappa)}\,
(l_{r}+C_{rs}l^{\dag}_{s})v_{r}(\kappa)\quad\text{and}\quad
\psi^{r}_{\kappa}
=-\sum_{r,s\geq \tfrac12}\frac{1}{\sqrt{2}\tilde{f}(\kappa)}\,
(r_{r}-C_{rs}r^{\dag}_{s})v_{r}(\kappa).
\end{gather}
This gives a precise relationship between the half-string
fermionic variables and the continuous
Moyal basis.

\section{Superghost vertex as Moyal kernel}
\label{sec:ghost-ver}
\setcounter{equation}{0}

The ghost vertex for the bosonic string was identified with
the Moyal product in the bosonized form in \cite{0204260}, \cite{0207174}
and in non-bosonized form in the Siegel gauge in \cite{0205107}.
The most direct way for an identification of superghost vertex with
the Moyal product will be to use the bosonized form of superghosts
and picture changing operator $Y_{-2}=Y(i)Y(-i)$.
However, in this section we use the non-bosonized language
to give a diagonalization of the minus one picture three-string
vertex,
which can be identified
with the continuous Moyal product in the manner analogous
to the one which was used in the Neveu-Schwarz matter case.
Similarly one can define the three-string vertex
in picture minus two, so that the string fields are in picture
zero. If all three vacua in the three-string
vertex are in picture minus two then we can
define the continuous variables making a shift in creation
and annihilation operators that corresponds to a shift
of all vacua from one picture to another.
Such diagonalization is alike the one
presented in this section but more cumbersome.

\subsection{Diagonalization of vertex}

In this subsection we introduce the continuous oscillators
and diagonalize the three-string superghost
vertex in the minus one picture.
The Neveu-Schwarz superghost
oscillators satisfy the following commutation relation
and hermitian conjugation properties
\begin{gather}
[\gamma_{r},\beta_{s}]=\delta_{r+s,0},\qquad
\gamma_{r}^{\dag}=\gamma_{-r},\quad \beta^{\dag}_{s}=-\beta_{-s}.
\label{commut-betagamma}
\end{gather}
The minus one picture vacuum is defined so that
\begin{gather}
\gamma_{r}|-1\rangle=0,\quad r\geq\tfrac12,\qquad
\beta_{s}|-1\rangle=0,\quad s\geq\tfrac12,
\end{gather}
and $\langle -1|-1\rangle=1$.
For this vacuum the BPZ conjugation is (bpz$^2=1$)
\begin{subequations}
\begin{gather}
\text{bpz}(\gamma_{-r})=(-1)^{-r-\tfrac12}\gamma_{r},\quad r\geq\tfrac12,\qquad
\text{bpz}(\gamma_{r})=(-1)^{r+\tfrac12}\gamma_{-r},\quad r\geq\tfrac12,\\
\text{bpz}(\beta_{-s})=(-1)^{-s+\tfrac32}\beta_{s},\quad s\geq\tfrac12,\qquad
\text{bpz}(\beta_{s})=(-1)^{s-\tfrac32}\beta_{-s},\quad s\geq\tfrac12.
\end{gather}
\label{beta-gamma-minus-bpz}
\end{subequations}
Let us define oscillators in the continuous basis as follows
\begin{subequations}
\begin{gather}
\tilde{\beta}^{\dag}_{\kappa}
=\sum_{r\geq\tfrac12}v_{r}(\kappa)\beta_{-r},\qquad
\tilde{\beta}_{\kappa}
=-\sum_{r\geq\tfrac12}v_{r}(\kappa)\beta_{r},\\
\tilde{\gamma}^{\dag}_{\kappa}
=\sum_{r\geq\tfrac12}v_{r}(\kappa)\gamma_{-r},\qquad
\tilde{\gamma}_{\kappa}
=\sum_{r\geq\tfrac12}v_{r}(\kappa)\gamma_{r}.
\end{gather}
\end{subequations}
The commutation relations \eqref{commut-betagamma}
in the continuous basis become
\begin{gather}
[\tilde{\gamma}_{\kappa},\tilde{\beta}^{\dag}_{\kappa'}]=
[\tilde{\beta}_{\kappa},\tilde{\gamma}^{\dag}_{\kappa'}]
=\delta(\kappa - \kappa').
\end{gather}
The twist operator $C$ acts in the continuous basis as follows:
$C\tilde{\beta}^{\dag}_{\kappa}=-\tilde{\beta}^{\dag}_{-\kappa}$
and $C\tilde{\gamma}^{\dag}_{\kappa}=-\tilde{\gamma}^{\dag}_{-\kappa}$.
Now introduce the even and odd under $C$ conjugation oscillators
\begin{subequations}
\begin{gather}
\beta^{\dag}_{\kappa,e}
=\frac{1}{\sqrt{2}}(
\tilde{\beta}^{\dag}_{\kappa}+C\tilde{\beta}^{\dag}_{\kappa}),\qquad
\beta^{\dag}_{\kappa,o}
=\frac{1}{\sqrt{2}}(
\tilde{\beta}^{\dag}_{\kappa}-C\tilde{\beta}^{\dag}_{\kappa}),\\
\gamma^{\dag}_{\kappa,e}
=\frac{1}{\sqrt{2}}(
\tilde{\gamma}^{\dag}_{\kappa}+C\tilde{\gamma}^{\dag}_{\kappa}),\qquad
\gamma^{\dag}_{\kappa,o}
=\frac{1}{\sqrt{2}}(
\tilde{\gamma}^{\dag}_{\kappa}-C\tilde{\gamma}^{\dag}_{\kappa}).
\end{gather}
\end{subequations}
We use the following
notations: $\beta_{\kappa,\alpha}^{\dag}=(\beta_{\kappa,e}^{\dag},\beta_{\kappa,o}^{\dag})$
and $\gamma_{\kappa,\alpha}^{\dag}=(\gamma_{\kappa,e}^{\dag},\gamma_{\kappa,o}^{\dag})$.
Using \eqref{beta-gamma-minus-bpz}
one finds the following BPZ conjugation
of the continuous oscillators
\begin{gather}
\text{bpz}(\beta^{\dag}_{\kappa,\alpha})=-c_{\alpha\beta}
\beta_{\kappa,\beta},\qquad
\text{bpz}(\gamma^{\dag}_{\kappa,\alpha})=c_{\alpha\beta}
\gamma_{\kappa,\beta}.
\end{gather}
The Neveu-Schwarz superghost three-string vertex in the minus one picture
in discrete and continuous basis is given by
\begin{gather}
|\tilde{V}_{3}\rangle_{123}
=\exp\Bigl[\sum_{a,b=1}^{3}\sum_{r,s\geq
\tfrac12} \beta_{-r}^{a}(C\tilde{M}^{ab})_{rs}\gamma^{b}_{-s}\Bigr]
|-1\rangle_{123};\\
|\tilde{V}_{3}\rangle_{123}
=\exp\Bigl[\int_{0}^{\infty}d\kappa\;
\beta_{\kappa,\alpha}^{\dag\,a}
\tilde{V}^{ab}_{\kappa,\alpha\beta}\gamma_{\kappa,\beta}^{\dag\,b}\Bigr]
|-1\rangle_{123}.
\label{diag-vert-ghost}
\end{gather}
Here $\tilde{V}^{ab}_{\kappa,\alpha\beta}$ is $6\times 6$ matrix of the form
\begin{gather}
\tilde{V}^{ab}_{\kappa,\alpha\beta}=
\tilde{\mu}(\kappa)\,\epsilon_{\alpha\beta}\otimes\delta^{ab}
+\tilde{\mu}_a(\kappa)\,c_{\alpha\beta}\otimes\chi^{ab}
+\tilde{\mu}_s(\kappa)\,\epsilon_{\alpha\beta}\otimes\varepsilon^{ab};
\end{gather}
and
\begin{gather}
\tilde{\mu}\equiv\tilde{\mu}^{11}
=-\tilde{\tau}\frac{1-\tilde{\tau}^{2}}{1+3\tilde{\tau}^2},\quad
\tilde{\mu}_{s}\equiv\frac12(\tilde{\mu}^{12}+\tilde{\mu}^{21})
=\tilde{\tau}\frac{1+\tilde{\tau}^{2}}{1+3\tilde{\tau}^2},\quad
\tilde{\mu}_{a}\equiv\frac12(\tilde{\mu}^{12}-\tilde{\mu}^{21})
=-\frac{1+\tilde{\tau}^{2}}{1+3\tilde{\tau}^2},
\end{gather}
where $\tilde{\tau}(\kappa)=\coth\left(\frac{\pi\kappa}{4}\right)$.

\subsection{Identification of Moyal structures}

In this subsection we introduce the coordinate
representation in which the three-string vertex
becomes the Moyal kernel.
We find the partition of unity and identify
the Moyal structures.
Introduce the continuous coordinate representation as follows
\begin{gather}
\langle \eta|=\langle-1|
\exp\Bigl[\int_{0}^{\infty}d\kappa\;
(\frac12\,\tilde{f}(\kappa)\eta^{\beta}_{\kappa,\alpha}
\,\epsilon_{\alpha\beta}\,\eta^{\gamma}_{\kappa,\beta}
+\eta^{\gamma}_{\kappa,\alpha}
\,c_{\alpha\beta}\,\beta_{\kappa,\beta}
-\eta^{\beta}_{\kappa,\alpha}
\,c_{\alpha\beta}\,\gamma_{\kappa,\beta}
+\tilde{\tau}(\kappa)
\beta_{\kappa,\alpha}\epsilon_{\alpha\beta}
\gamma_{\kappa,\beta})\Bigr],
\end{gather}
where $\eta^{\beta}_{\kappa,\alpha}
=(\eta^{\beta}_{\kappa,e},\eta^{\beta}_{\kappa,o})$
and $\eta^{\gamma}_{\kappa,\alpha}
=(\eta^{\gamma}_{\kappa,e},\eta^{\gamma}_{\kappa,o})$.
The BPZ conjugated state
$\text{bpz}(\langle\tilde{\eta}|)\equiv|\tilde{\eta}\rangle$ is given by
\begin{gather}
|\eta\rangle=
\exp\Bigl[\int_{0}^{\infty}d\kappa\;
(\frac12\,\tilde{f}(\kappa)\eta^{\beta}_{\kappa,\alpha}
\,\epsilon_{\alpha\beta}\,\eta^{\gamma}_{\kappa,\beta}
-\eta^{\gamma}_{\kappa,\alpha}
\,\delta_{\alpha\beta}\,\beta^{\dag}_{\kappa,\beta}
-\eta^{\beta}_{\kappa,\alpha}
\,\delta_{\alpha\beta}\,\gamma^{\dag}_{\kappa,\beta}
+\tilde{\tau}(\kappa)
\beta^{\dag}_{\kappa,\alpha}\epsilon_{\alpha\beta}
\gamma^{\dag}_{\kappa,\beta})\Bigr]|-1\rangle.
\label{coor-repr-right-betagamma}
\end{gather}
Let us rewrite this coordinate representation
in discrete basis.
Introduce the discrete variables
\begin{subequations}
\begin{align}
\tilde{\eta}^{\beta}_{\kappa}&=\sum_{r\geq\tfrac12}v_{r}(\kappa)\eta^{\beta}_{r},
&\tilde{\eta}^{\gamma}_{\kappa}&=\sum_{r\geq\tfrac12}v_{r}(\kappa)\eta^{\gamma}_{r},\\
\eta^{\beta}_{\kappa,e}&
=\sqrt{2}\sum_{r_e}v_{r_e}(\kappa)\eta^{\beta}_{r_e},
&\eta^{\gamma}_{\kappa,e}&
=\sqrt{2}\sum_{r_e}v_{r_e}(\kappa)\eta^{\gamma}_{r_e},\\
\eta^{\beta}_{\kappa,o}&
=\sqrt{2}\sum_{r_o}v_{r_o}(\kappa)\eta^{\beta}_{r_o},
&\eta^{\gamma}_{\kappa,o}&
=\sqrt{2}\sum_{r_o}v_{r_o}(\kappa)\eta^{\gamma}_{r_o}.
\end{align}
\end{subequations}
In terms of these discrete variables the coordinate representation
\eqref{coor-repr-right-betagamma} takes the form
\begin{gather}
|\eta\rangle=\exp\Bigl[
\,\frac12\,\sum_{r,s\geq\tfrac12}\eta^{\beta}_r\,\tilde{F}_{rs}\,\eta^{\gamma}_s
-\sum_{r\geq\tfrac12}\eta^{\gamma}_r\,\beta_{-r}
-\sum_{r\geq\tfrac12}\eta^{\beta}_r\,\gamma_{-r}
+\sum_{r,s\geq\tfrac12}\beta_{-r}\,\widetilde{I}_{rs}\,\gamma_{-s}
\Bigr]|-1\rangle.
\end{gather}
Notice that $|\eta\rangle$
has an interpretation of the coherent state over
the identity surface state in the picture minus one.
One finds in the continuous and discrete basis
\begin{subequations}
\begin{gather}
\langle \eta|\lambda\rangle
=\tilde{\Nc}_{I}
\exp\Bigl[\int_{0}^{\infty}d\kappa\; \frac{1}{\tilde{\theta}(\kappa)}\,
(\eta^{\beta}_{\kappa,\alpha}\,c_{\alpha\beta}\,
\lambda^{\gamma}_{\kappa,\beta}
-\eta^{\gamma}_{\kappa,\alpha}\,c_{\alpha\beta}\,
\lambda^{\beta}_{\kappa,\beta})\Bigr],\\
\langle \eta|\lambda\rangle=\tilde{\Nc}_{I}
\exp\Bigl[\,\frac12\sum_{r,s\geq\tfrac12}
(\eta^{\beta}_{r}\,((1+F)C)_{rs}\,\lambda^{\gamma}_{s}
-\eta^{\gamma}_{r}\,((1+F)C)_{rs}\,\lambda^{\beta}_{s}
)\Bigr],
\end{gather}
\end{subequations}
where $\tilde{\theta}(\kappa)=1+\tilde{\tau}^2(\kappa)$
and $\tilde{\Nc}_{I}=\det(1+(C\tilde{I})^2)^{-1}$.
The partitions of unity in the continuous and discrete basis are given by
\begin{subequations}
\begin{gather}
1=\tilde{\Nc}_{I}\int \mathbf{d}\eta^{\beta}\,\mathbf{d}\eta^{\gamma}\,
 \mathbf{d}\lambda^{\beta}\,\mathbf{d}\lambda^{\gamma}\,
\exp\Bigl[-\int_{0}^{\infty}d\kappa\,\frac{1}{\tilde{\theta}(\kappa)}\,
(\eta^{\beta}_{\kappa,\alpha}\,c_{\alpha\beta}\,
\lambda^{\gamma}_{\kappa,\beta}
-\eta^{\gamma}_{\kappa,\alpha}\,c_{\alpha\beta}\,
\lambda^{\beta}_{\kappa,\beta})\Bigr]
|\eta\rangle\langle\lambda|;\\
1=\tilde{\Nc}_{I}\int \mathbf{d}\eta^{\beta}\,\mathbf{d}\eta^{\gamma}\,
\mathbf{d}\lambda^{\beta}\,\mathbf{d}\lambda^{\gamma}\,
\exp\Bigl[\,-\frac12\sum_{r,s\geq\tfrac12}
(\eta^{\beta}_{r}\,((1+F)C)_{rs}\,\lambda^{\gamma}_{s}
-\eta^{\gamma}_{r}\,((1+F)C)_{rs}\,\lambda^{\beta}_{s}
)\Bigr]|\eta\rangle\langle\lambda|.
\end{gather}
\label{edinitsa-betagamma}
\end{subequations}
Here
\begin{gather}
\mathbf{d}\eta^{\beta}\equiv\prod_{r\geq\tfrac12}
d\eta^{\beta}_{r}\,,\qquad
\mathbf{d}\eta^{\gamma}\equiv\prod_{r\geq\tfrac12}
d\eta^{\gamma}_{r}.
\end{gather}

The functional corresponding to string field $|\Psi\rangle$
in the minus one picture is given by
$\Psi(\eta)\equiv
\langle\eta|\Psi\rangle=\langle\Psi|\eta\rangle$.
Inserting two partitions of unity
\eqref{edinitsa-betagamma} in the three-string superghost star product
\begin{gather}
|\Psi^{1}*\Psi^{2}\rangle_{3}
={}_{1}\langle\Psi^1|{}_{2}\langle\Psi^{2}|\tilde{V}_{3}\rangle_{123}
\end{gather}
one finally finds
\begin{gather}
(\Psi^{1}*\Psi^{2})(\eta^{3})
=\int
\mathbf{d}\eta^{\beta\,1}\,\mathbf{d}\eta^{\gamma\,1}\,
\mathbf{d}\eta^{\beta\,2}\,\mathbf{d}\eta^{\gamma\,2}\,
\Psi^{1}(\eta^1)
\Psi^{2}(-\eta^2)
K(\eta^1,\eta^2,\eta^3).
\end{gather}
Here the kernel
corresponding to the three-string vertex \eqref{diag-vert-ghost}
is given by
\begin{gather}
K(\eta^1,\eta^2,\eta^3)\equiv
{}_{1}\langle \eta^1|
{}_{2}\langle \eta^2|
{}_{3}\langle \eta^3|\tilde{V}_{3}\rangle_{123}
=\tilde{\Nc}_{K}\exp\Bigl[\int_{0}^{\infty}d\kappa\;
\frac{1}{\tilde{\theta}(\kappa)}\,
\eta^{\beta\,a}_{\kappa,\alpha}
\,c_{\alpha\beta}\otimes \chi^{ab}\,
\eta^{\gamma\,b}_{\kappa,\beta}\Bigr],
\end{gather}
 and
\begin{gather}
\tilde{\Nc}_{K}=\det\left(
\tfrac14(1+F)^{2}(2-F)\right).
\end{gather}

\section*{Acknowledgments}
We would like to thank D. Belov, A. Koshelev and I. Volovich for
useful discussions. This work was supported in part by RFBR grant
02-01-00695  and RFBR grant for leading scientific schools and  by
INTAS grant 99-0590.

\appendix
\section*{Appendices}
\addcontentsline{toc}{section}{Appendices}
\renewcommand {\theequation}{\thesection.\arabic{equation}}

\section{The three-string matter vertex}
\setcounter{equation}{0}

The Neveu-Schwarz three-string matter
vertex and identity Neumann matrices are expressed
in terms of the
real symmetric matrices
$F$ and $C\tilde{F}$ as \cite{GJ3}
\begin{gather}
M^{11}=\frac{FC\tilde{F}}{(1-F)(2+F)},\;
M^{12}=\frac{C\tilde{F}+(1-F)}{(1-F)(2+F)},\;
M^{21}=\frac{C\tilde{F}-(1-F)}{(1-F)(2+F)},\;
CI=\frac{C\tilde{F}}{1-F}.
\end{gather}
They satisfy the following cyclic property
\begin{gather}
M^{a+1\,b+1}=M^{a\,b},\;\;
\forall \;\;a,b,a+1,b+1(\text{ mod } 3).
\end{gather}
These matrices are commuting and have the same as $F$ and $C\tilde{F}$
set of eigenvectors
\begin{gather}
\sum_{s\geq\tfrac12}M^{ab}_{rs}v_{s}(\kappa)
=\mu^{ab}(\kappa)v_{r}(\kappa),\qquad
\sum_{s\geq\tfrac12}(CI)_{rs}v_{s}(\kappa)
=-\tau(\kappa)v_{r}(\kappa),
\label{eigen-eq}
\end{gather}
the eigenvalues \eqref{mu-tau} are given in \cite{0112231}.
Note also that $\mu^{ab}(-\kappa)=-\mu^{ba}(\kappa)$.

The hermitian matrices $F_{rs}$ and
$(C\tilde{F})_{rs}$ are defined by
\begin{gather}
F_{rs}=-\frac{2}{\pi}\frac{\imath^{r-s}}{r+s},\quad
r=s\text{ mod }2,\qquad
\tilde{F}_{rs}=-\frac{2}{\pi}\frac{\imath^{r+s}}{r-s},\quad
r=s+1\text{ mod }2.
\end{gather}
They have the following properties
\begin{gather}
F^{2}-\tilde{F}^{2}=1,\quad [F,\tilde{F}]=0,\quad
[F,C]=0,\quad F^{T}=F,\quad \{\tilde{F},C\}=0,\quad
\tilde{F}^{T}=-\tilde{F}.
\end{gather}

\section{The three-string ghost vertex}
\setcounter{equation}{0}

The three-string Neveu-Schwarz superghost vertex
in the minus one picture has the following Neumann matrices
\begin{gather}
\tilde{M}^{11}=\frac{FC\tilde{F}}{(1+F)(2-F)},\quad
\tilde{M}^{12}=\frac{-C\tilde{F}-(1+F)}{(1+F)(2-F)},\quad
\tilde{M}^{21}=\frac{-C\tilde{F}+(1+F)}{(1+F)(2-F)},\quad
C\tilde{I}=-\frac{C\tilde{F}}{1+F}.
\end{gather}
They have the same set of eigenvectors as the matter vertex matrices.
The property $\tilde{\mu}^{ab}(-\kappa)=-\tilde{\mu}^{ba}(\kappa)$ is obeyed.

\end{document}